\documentclass[showpacs,aps,preprint]{revtex4}
\usepackage{amsmath}
\usepackage{amsfonts}
\usepackage{amssymb}
\usepackage{amsthm}
\usepackage{graphicx}
\usepackage{color}

\begin{document}


\title{Light emitting diodes of inverse spin valves}
\author{X. R. Wang}
\affiliation{ Department of Physics, The Hong Kong University 
of Science and Technology, Clear Water Bay, Hong Kong, China}
\date{\today}

\begin{abstract}
We propose making light emitting diodes out of inverse
spin valves. The proposed diodes rely on the spin-dependent 
electron transport of inverse spin valves that are layered 
structures of a ferromagnetic half-metal sandwiched between 
two non-magnetic metals. Under a bias, a giant spin-dependent 
chemical potential difference between spin-up and spin-down 
electrons is created. Thus, the inverse spin valves are 
possible to emit light when electrons in higher chemical 
potential go to the lower chemical potential. 
The advantages of this type of light emitting diodes 
include tunableness and less demand on materials.
\end{abstract}
\keywords{LED, Spintronics, Polarized light}
\pacs{} \maketitle

Most light emitting diodes (LEDs) are made of either 
non-magnetic semiconductors or organic semiconductors. 
The working principle of an LED is to create a situation 
where electrons occupy higher energy levels and some of 
the lower energy levels are unoccupied so that electrons 
can move to those lower energy levels by emitting photons. 
One of the popular ways to achieve this in the conventional 
LEDs is by a p-n junction where electrons are injected into 
the conduction band on one side and holes into the valence 
band on the other side (of the junction) when the LED is 
electrically biased in the forward direction of the junction. 
As long as an electron and a hole are spatially not too far 
from each other, they can radiatively recombine and emit 
incoherent narrow-spectrum light. 
Thus, the emitted light depends largely on the energy 
difference between electrons and holes. This is also why proper 
electronic structure are normally required for LED materials. 
The tunableness of the light frequency depends on a particular 
energy spectrum of the material used. On the other hand, 
all materials can emit light if they are in excited states. 
The question that we would like to ask is whether it is possible 
to make an LED out of materials that are normally not for LEDs. 
In this letter, we propose an inverse spin valve device in 
which the chemical potentials (Fermi levels) of spin-up (SU) and 
spin-down (SD) electrons split when an external bias is applied.    
The degree of the split is at the magnitude of applied bias, 
and is due to the spin-dependent electron transport.
Thus the device can emit tunable light under an external bias. 
Interestingly enough, this inverse spin valve is made out of 
both non-magnetic and magnetic metals. Traditionally, one will 
not relate LEDs with magnetism where one may be interested in 
the magnetization reversal\cite{xrw}. Furthermore, the emitted 
light is not sensitive to the particular materials used, and 
thus it is an LED out of any material. 

A conventional spin valve is a layered structure of a 
non-magnetic spacer sandwiched between two ferromagnetic metals. 
An inverse spin valve is a layered structure with a 
ferromagnet sandwiched between two non-magnetic metals. 
As illustrated in Fig. 1a, $M1$ and $M2$ are two non-magnetic 
metals. To maximize the spin-related chemical potential split, 
the ferromagnetic spacer is chosen to be a half-metal (HM) 
that acts as a conductor to electrons of one spin orientation 
(spin-up), but as an insulator to those of the opposite 
orientation (spin-down). Then only spin-up electrons can pass 
through the half-metal when an external bias is applied on 
the inverse spin valve. Spin-down electrons will be blockaded 
from the flow through the spacer. Similar to a giant 
magnetoresistance\cite{Fert,Grun,Parkin} or tunneling 
magnetoresistance\cite{Miyazaki,Moodera} device, electron 
transport of an inverse spin valve is spin-dependent. 

The spin-dependent electron transport of an inverse spin 
valve leads to a spin-dependent Fermi levels in non-magnetic 
metals near the metal-ferromagnet interfaces under a bias. 
Consider one inverse spin valve connected to a battery of 
voltage $V$. Let us assume that the electron chemical 
potential in $M1$ is initially moved up by $eV$ while that 
of $M2$ is kept unchanged\cite{sdwang}. The electron flow 
diagram is shown in Fig. 1b. $M1$ and $M2$ each has two 
electron reservoirs. One is for SU electrons, and the other 
is for SD electrons (denoted by rectangular boxes). 
SU electrons in $M1$ can flow into the empty SU electronic 
states in $M2$ via the empty SU electronic states in the 
half-metal, generating a current $I$ from $M1$ to $M2$. 
There is no current between the SD-electron reservoir of $M1$ 
and SD-electron reservoir of $M2$ due to the spin blockade of 
half-metal. Same amount electrons will be pumped back from $M2$ 
to $M1$ by the battery to keep the electron neutrality in $M1$ 
and $M2$. However, a battery does not distinguish electron 
spin, and it pumps equal amount of SU and SD electrons. 
In other words, SU electrons flow out of $M1$ and into $M2$.
In the meanwhile, an equal amount of electrons with half of 
them in the SU state and the other half in the SD state are 
drawn out of $M2$ and are supplied into $M1$ by a battery. 
As a result, $M1$ accumulates more SD electrons, and $M2$ 
accumulates more SU electrons. Thus, the chemical potential of 
the SD electrons is higher than that of SU electrons in $M1$. 
Vice versa, the Fermi level of SU electrons is higher than 
that of SD electrons in $M2$. Fig. 1c is an illustration 
of the Fermi levels (denoted by the dash lines) of SU and SD 
electrons in non-magnetic metals and half-metal. 
The electron density of states (DOSs) in the figure for 
non-magnetic metals and half-metal is just a sketch. 
$\Delta\mu_1$ and $\Delta\mu_2$ are the chemical potential 
differences between SU and SD electrons in $M1$ and $M2$, 
respectively. $\Delta V$ is the chemical potential difference 
between the SU electrons in $M1$ and the SU electrons in $M2$. 

SD electrons in $M1$ can only go to $M2$ by first flipping
their spins and changing to SU electrons. The only supply of
the SD electrons to $M2$ is from the conversion of the SU 
electrons in $M2$ through spin flipping. As shown in Fig. 1b, 
these lead to the internal currents $I_1$ in $M1$ and $I_2$ 
in $M2$ between SD-electron to SU-electron reservoirs. 
Assume the spin flip occurs only near non-magnetic-magnetic 
interfaces within a width of spin diffusion length $\xi_1$ in 
$M1$ and $\xi_2$ in $M2$. This is justified because $\xi_i$ 
($i=1,2$) is the length scale over which a chemical potential 
difference can maintain, and the conversion rate of SU and SD 
electrons from each other is the same when both SU and SD 
electrons have the same Fermi levels (chemical potentials). 
Let $\tau_1$ and $\tau_2$ be the spin flipping time 
(spin-relaxation time $T_1$\cite{t1t2}) in $M1$ and $M2$,
corresponding to the flipping rate of $1/\tau_1$ and $1/\tau
_2$, respectively. The conversion rate from the SD electrons to 
the SU electrons in $M1$ is the product of the excess the SD 
electrons $n_1 \Delta\mu_1 \xi_1 A$ and the single electron
flipping rate. Thus $I_1$ is 
\begin{equation}\label{current1}
I_1=\frac{n_1\Delta \mu_1e\xi_1 A }{\tau_1},
\end{equation}
where $n_1$ is the density of states of the SD electrons in 
$M1$ at the Fermi level. $A$ is the cross section of $M1$. 
Similarly, the current due to the conversion of the SU 
electrons to the SD electrons in $M2$ is
\begin{equation}\label{current2}
I_2=\frac{n_2\Delta \mu_2e\xi_2 A }{\tau_2}.
\end{equation}

The current $I$ from $M1$ to $M2$ is, neglecting the tunneling
by the SD electrons through the half-metal and assuming a
resistance $R$ for the SU electrons,
\begin{equation}\label{current}
I=\frac{\Delta V}{R}. 
\end{equation}
At the steady state, there is no net electron build up 
anywhere in the circuit. Since the current through the 
battery is unpolarized, half of the current is made up by 
the SU electrons and the other half is from the SD electrons. 
Balance conditions and external constraint require 
\begin{align}\label{steady}
&I_1=I/2 ,\nonumber\\
&I_2=I/2 , \\
&\Delta\mu_1/e+\Delta\mu_2/e+\Delta V=V. \nonumber 
\end{align}
Solving Eqs. 1-4, the spin-dependent chemical potential 
differences $\Delta\mu_1$ and $\Delta\mu_2$ are 
\begin{align}\label{split}
&\Delta\mu_1=\frac{(eV)\tau_1/(n_1e^2\xi_1A)}
{2R+\tau_1/(n_1e^2\xi_1A)+\tau_2/(n_2e^2\xi_2A)},\nonumber \\ 
&\Delta\mu_2=\frac{(eV)\tau_2/(n_2e^2\xi_2A)}
{2R+\tau_1/(n_1e^2\xi_1A)+\tau_2/(n_2e^2\xi_2A)}.
\end{align}
It is interesting to see that the largest chemical potential 
splits occur at $R=0$, a short circuit for spin-up (SU) electrons! 
The half-metal may also be replaced by an ordinary ferromagnet. 
In this case, SD electrons in $M1$ can also flow directly into 
$M2$. As long as there is a spin-dependent electron 
transportation, the spin-related chemical potential differences 
in $M1$ and $M2$ always exist but their values will be reduced 
by a factor of $(1-R/R')$\cite{xrw1}, where $R'$ is the resistance 
of the Ferromagnet for the SD electrons (minority carriers).  

Results of Eq. \eqref{split} means the population inversion of 
the electrons in the electrically biased inverse spin valves.
As shown in Fig. 2a for $M1$, SU electronic states below their 
Fermi level are fully occupied (at zero temperature) while the SD 
electronic states in the energy range of $\Delta\mu_1$ between 
SU electron Fermi level and SD electron Fermi level are empty. 
Thus an SU electron can go to a lower empty SD electronic state 
and emit a photon. One can then apply standard theory of LED to 
the inverse spin valve device to make an LED or even a laser by 
incorporating the device with a cavity\cite{laser}. However, 
unlike the usual light source of atoms and semiconductors where 
spin does not involve in the electron transitions, the light 
source here is spin resolved. For example, assume the magnetization 
of the half-metal is along the +z-direction. The electrons in 
higher energy levels of $M1$ are in spin-up (+z-direction) states. 
They can only go to spin-down states. Due to the angular momentum 
conservation, the spin of the emitted photon must be along the 
+z-direction. In other words, the light out of an inverse spin 
valve is polarized. Its polarization depends on the light 
propagation direction and magnetization of the half-metal. 
As shown in Fig. 2b, the light emitted in the +z-direction 
is right-hand circularly polarized light, and left-hand 
circularly polarized light in the -z-direction. 
Because of the two-component nature of photons (angular 
momentums must be parallel to light propagation direction), 
no-light can emit in the direction perpendicular to the 
magnetization of the half-metal (xy-plane). 

In comparison with the usual semiconductor LEDs, there are a few 
nice features about the proposed LED. 1) Photons have a well 
defined polarization because both occupied and unoccupied states 
have well defined spins. 2) Two electronic states involved in a 
transition are in the same physical locations so that the 
oscillator strength (transition matrix element) should be large.  
3) Since the spin-related chemical potential split is due to 
the spin-dependent electron transport (electrons of one spin  
orientation is blockaded from the flow in the inverse spin valve), 
the population inversion is not very sensitive to the temperature. 
Thus, there is no reason to prevent the current LED device 
to function at room temperature. 4) The population inversion is 
not very sensitive to the detail electronic structure, thus the 
physics is very robust, and one can in principle use any 
conducting materials, magnetic or non-magnetic and organic or 
inorganic, to make LEDs. 5) The spin-dependent chemical potential 
difference is controlled by an external bias, thus the light 
frequency can be electrically tuned over very broad range. 

Zero temperature is assumed in Eqs. \eqref{current1} and 
\eqref{current2}. For a finite temperature, electron 
distribution is no longer described by a step-function, and 
Eqs. \eqref{current1} and \eqref{current2} should be modified. 
Also, the spatial variation of the Fermi level is neglected 
in the present work. In principle, Fermi levels of both 
spin-up and spin-down electrons are position-dependent due to 
electron diffusion, spatial distribution of electrons, and 
spin flipping. The position-dependence of the Fermi levels 
will also modify Eqs. \eqref{current1} and \eqref{current2}. 
For practical applications, it is interesting to work out a 
more careful analysis by taking into account the electron 
distribution in both energy and space although one should 
not anticipate any change in physics. 

The working principle of currently proposed LED device is very 
different from the traditional ones. Unlike all previous LEDs 
where electron spin degrees of freedom were not used, the LED 
devices here is largely based on manipulating electron spins.  
The population inversion does not occur between states of the 
same spin, but for opposite spin instead. 

In conclusion, we propose a very robust LED device made of 
almost any material. Unlike the usual LEDs that rely on detail 
energy spectrum of the material, the proposed technology uses 
electron spin blockage to create a population inversion of 
electrons. 

The author would like to thank Prof. Yongli Gao and Prof. John 
Xiao for the useful discussions. This work is supported by 
Hong Kong RGC/CERG grants (\# 603007 and SBI07/08.SC09).

\newpage
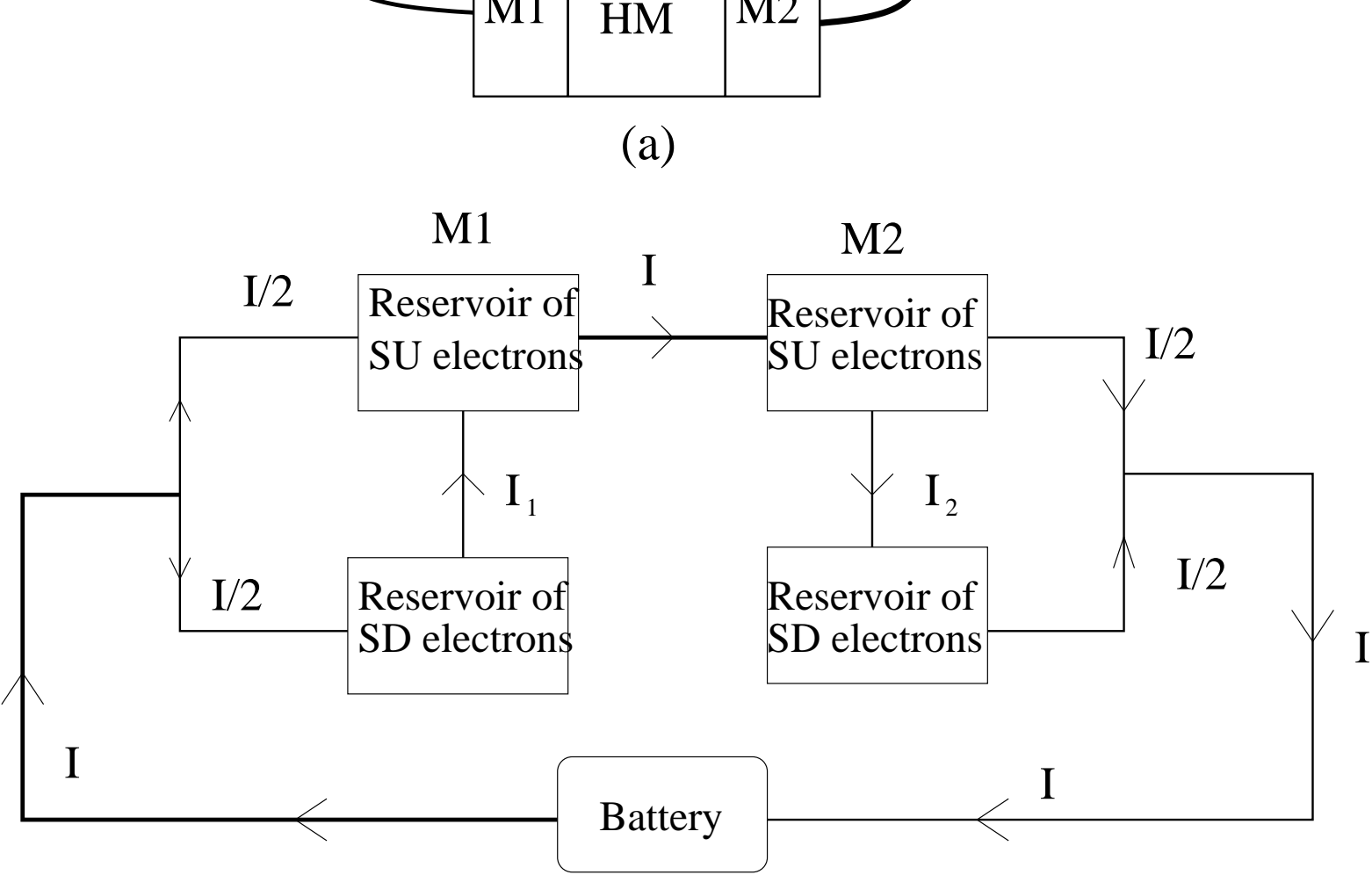
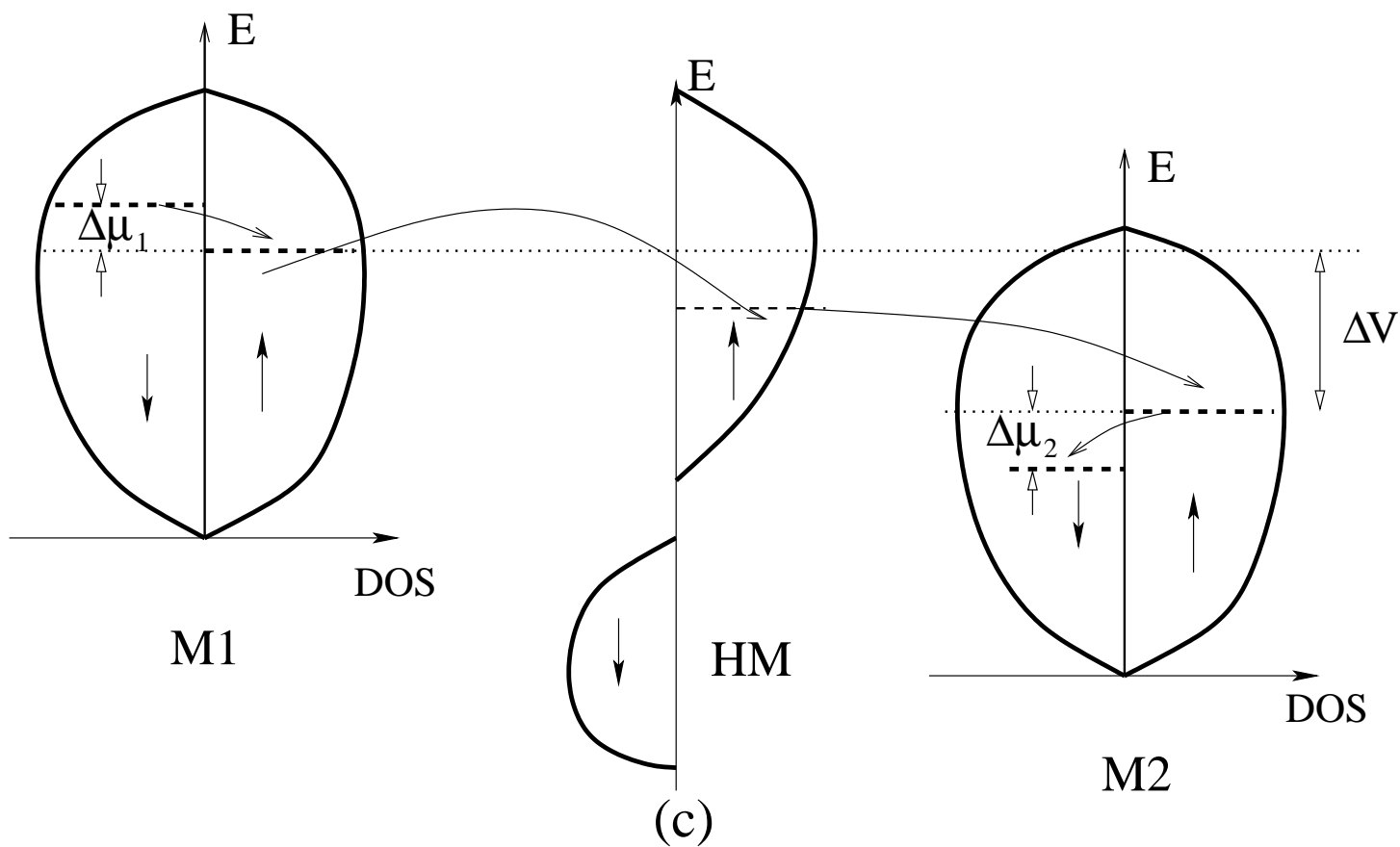
\begin{figure}[htbp]
\caption{\label{fig1} (a) Schematic illustration of the 
inverse spin valve. $M1$ and $M2$ are non-magnetic metals. 
HM is a ferromagnetic half metal. $V$ is the applied bias. 
(b) Electron flow from and into spin-up and spin-down states 
in $M1$ and $M2$. At the steady state, current flowing 
into any reservoir should be equal to those flowing out. 
(c) Relative chemical potentials of spin-up and spin-down 
electrons in non-magnetic metals and half-metal. 
The curved arrows indicate the electron flow. Dash lines 
indicate the chemical potential levels of spin-up and spin 
down electrons in half-metal and non-magnetic metals. 
$\Delta\mu_1$ and $\Delta\mu_2$ are the chemical 
potential splits between spin-up and spin-down electrons 
at the left and right metal-ferromagnet interfaces. 
$\Delta V$ is the effective bias on the half-metal.}
\end{figure}
\begin{figure}[htbp]
\caption{\label{fig2} (a) Schematic illustration of light 
emitting process in $M1$. SU electrons near their Fermi level 
can jump to the lower empty SD electronic level, and emit 
photons of well-defined polarization. 
(b) A sketch of possible LED out of inverse spin valve. 
If the magnetization of the half-metal is along the z-axis, the 
light emitted by $M1$ along the +z-direction shall be right-hand 
circularly polarized. No light shall emit in the xy-plane. }
\end{figure}

\newpage
\begin{thebibliography}{}
\bibitem{xrw} X.R. Wang, and Z.Z. Sun, Phys. Rev. Lett.
{\bf 98}, 077201 (2007); Z.Z. Sun, and X.R. Wang, {\it ibid}
{\bf 97}, 077205 (2006); Phys. Rev. B {\bf 71}, 174430 (2005); 
{\it ibid} {\bf 73}, 092416 (2006); {\it ibid} {\bf 74}, 132401 
(2006); T. Moriyama, R. Cao, J.Q. Xiao, J. Lu, X.R. Wang, Q. Wen, 
and H.W. Zhang, Appl. Phys. Lett. {\bf 90}, 152503 (2007). 
\bibitem{Fert}  M.N. Baibich, J.M. Broto, A. Fert, F.N. Van 
Dau, F. Petroff, P. Etienne, G. Creuzet, A. Friederich, 
and J. Chazelas, Phys. Rev. Lett. {\bf 61}, 2427 (1988). 
\bibitem{Grun} G. Binach, P. Grunberg, F. Saurenbach, 
and W. Zinn, Phys. Rev. B {\bf 39}, 4828 (1989). 
\bibitem{Parkin} R.F.C. Farrow, C.H. Lee, and S.S.P. Parkin, 
IBM J. Res. Dev. {bf 34}, 903 (1990). 
\bibitem{Miyazaki} T. Miyazaki and N. Tezuka, J. Magn. Magn. 
Mater. {\bf 139}, L231 (1995).   
\bibitem{Moodera} J.S. Moodera et al., Phys. Rev. Lett. 
{\bf 74}, 3273 (1995).
\bibitem{sdwang} S.D. Wang,  Z.Z. Sun, N. Cue, H.Q. Xu, and 
X.R. Wang, Phys. Rev. B {\bf 65}, 125307 (2002).
\bibitem{t1t2} X.R. Wang, Y.S. Zheng, and S. Yin, Phys. Rev. B 
{\bf 72}, R121303 (2005).
\bibitem{xrw1} X.R. Wang, unpublished. 
\bibitem{laser} {\it Laser theory }, Hermann Haken, 
Springer-Verlag, 1983. 
\end{thebibliography}
\end{document}